# Graphene on Au-coated SiO$_x$ substrate: Its core-level photoelectron micro-spectroscopy study

Jhih-Wei Chen[1], Chiang-Lun Wang[1], Hung Wei Shiu[2], Chi-Yuan Lin[3], Chen-Shiung Chang[3], Forest Shih-Sen Chien[4], Chia-Hao Chen[2], Yi-Chun Chen[1], and Chung-Lin Wu[*,1,5]

[1]*Department of Physics, National Cheng Kung University, Tainan 70101, Taiwan*
[2]*National Synchrotron Radiation Research Center, Hsinchu 30076, Taiwan*
[3]*Department of Photonics and Institute of Electro-Optical Engineering, National Chiao Tung University, Hsinchu 30010, Taiwan*
[4]*Department of Physics, Tunghai University, Taichung 40704, Taiwan*
[5]*Advanced Optoelectronic Technology Center, National Cheng Kung University, Tainan 70101, Taiwan*

## Abstract

The core-level electronic structures of the exfoliated graphene sheets on a Au-coated SiO$_x$ substrate have been studied by synchrotron radiation photoelectron spectroscopy (SR-PES) on a micron-scale. The graphene was firstly demonstrated its visibility on the Au-coated SiO$_x$ substrate by micro-optical characterization, and then conducted into SR-PES study. Because of the elimination of charging effect, precise C 1$s$ core-level characterization clearly shows graphitic and contaminated carbon states of graphene. Different levels of Au-coating-induced *p*-type doping on single- and double-layer graphene sheets were also examined in the C 1$s$ core-level shift. The Au-coated SiO$_x$ substrate can be treated as a simple but high-throughput platform for *in situ* studying graphene under further hybridization by PES.

[*]Corresponding author; email address: clwuphys@mail.ncku.edu.tw

Graphene (a monolayer of graphite) is a unique low-dimensional system with fascinating electrical properties including linear electron energy dispersion and high room-temperature mobility,[1,2] which has been considered a prospective candidate for post-silicon electronics. Recent progress in pristine graphene preparation has mainly been on exfoliated graphene, which is fabricated by micromechanical cleavage of graphite and then transferred to certain substrates that render it visible. A Si substrate capped with a 300 nm insulating $SiO_x$ dielectric layer is frequently used for supporting graphene. In the gapless band structure of pristine graphene, the Fermi level coincides with its conical points, but graphene on $SiO_x$ substrate or metal surfaces would significantly shifts its band structure.[3,4] Besides the concerns on the influence of graphene bottom contacts, recent studies attempted to functionalize the graphene top surface by absorbing or incorporating various materials to modify its electronic structure, e.g., control band gap formation.[5] Therefore, through the band structure shift and core-level variation of graphene, how to consistently understand the coupling of graphene to its environment is interesting for fundamental studies and crucial for applying graphene to future devices.

With extensive analysis based on high-resolution core-level C 1*s* spectra, photoelectron spectroscopy (PES) is regarded as a powerful tool for graphene characterization. By characterizing the valence band and core-level spectra of graphene, PES provides direct information on hybridization with the environment, and determines graphene metallicity, defect density, and contamination.[6,7,8] However, the presence of the normally used insulating $SiO_x$ substrate complicates PES characterization. Although PES can measure graphene on a $SiO_x$ substrate, previous results remain unclear on local charging and screening effects.[9,10] In this letter, we demonstrate that a thin conducting Au film grown on $SiO_x$ can serve as a graphene-supporting substrate. This substrate shows similar optical microscope (OM) contrast as the pristine $SiO_x$ substrate does, and successfully performs micro-Raman investigations on the proposed system. The conducting Au thin film enables precise core-level PES studies on exfoliated single- (SLG), bilayer- (BLG) and multi-layer graphene (MLG) using local photoelectron image/spectroscopy on a micron scale. The graphitic $sp^2$ carbon and contaminated carbon states of graphene were precisely characterized in the core-level spectroscopy. Different levels of Au-coating-induced *p*-type doping of SLG and BLG were also examined by its core-level shifts.

The graphene samples were prepared by mechanical exfoliation on Si (100) substrate covered with 300 nm $SiO_x$ and 9 nm Au layer. The sputtered thin Au film is particularly suitable for PES measurements because Au is a bottom electrode for preventing local photoelectron charging, and does not obscure the optical interference that makes graphene visible to have instant optical inspection. The graphene sheets with different numbers of layers were identified using an optical microscope, and also confirmed by micro-Raman spectroscopy (μ-Raman) measurements before conducting localized PES studies. The localized synchrotron-radiation PES (SR-PES) studies were performed by the monochromatic (480 eV) soft X-ray at the scanning photoelectron microscopy (SPEM) end station at the National Synchrotron Radiation Research Center (NSRRC) in Hsinchu, Taiwan. The soft X-ray beam focused by the Fresnel zone plate and order-sorting aperture at the focal plane was about 100-200 nm in diameter. The emitted photoelectrons were synchronized collected by a multiple-channel hemispherical electron energy analyzer. While raster-scanning the sample, a two-dimensional distribution of that particular element can be mapped by setting the electron collecting energy window of the analyzer to a characteristic core-level emission. Based on the SPEM images, the focused beam can be moved to specific locations to perform high-resolution microscopic-area PES (μ-PES).

Figure 1 summarizes the optical image and spectroscopy results of the graphene sheets having different layer thicknesses on the Au-coated $SiO_x$ substrate. The graphene thicknesses were identified using an optical microscope, and then confirmed by μ-Raman measurements. Generally, to enhance the interference effect between the graphene and $SiO_x$ layer, a 300 nm-thick $SiO_x$ layer is optimal for white light illumination.[11] Because the Au layer was thin enough, the OM images shown in Fig.

1(a) demonstrated that optical contrast was sufficient for graphene identification where blue- and red-dashed lines enclose the SLG and BLG regions separately. The OM image contrast of graphene on the Au/SiO$_x$ substrate was quantitatively evaluated through the reflection spectrum (RS) shown in Fig. 1(b). Using Fresnel's model for the graphene/Au/SiO$_x$ tri-layer system on the Si substrate, the 9 nm Au thin film on SiO$_x$ substrate shows best fitting in the RS with the highest visibility of SLG under 550 nm wavelength visible light, and the validity of exfoliated graphene characterization is given. The μ-Raman characterized the quality and the number of layer of exfoliated graphene sheets that we identified with OM image. The strong Raman background signal from Au significantly influenced the graphene G band (SLG, BLG, and MLG) between 1100 to 1700 cm$^{-1}$; therefore, the 2D band was used here to identify the number of graphene layers. In Fig. 1(c), the Raman spectra of SLG, a very sharp (~30 cm$^{-1}$ in width) and symmetric 2D band at around 2634 cm$^{-1}$ is clearly present. In contrast, the 2D band of BLG and MLG are considerably broader (~59 cm$^{-1}$ and ~70 cm$^{-1}$ in width) and can be fitted by multi-peaks. According to the double resonance theory, the variation in the 2D band as thickness increases is attributed to the evolution of multi-layered graphene's electronic structure.[12]

After optically characterizing the graphene sheets on the Au/SiO$_x$ substrate, Fig. 2 (a) and (b) show the SPEM images, corresponding to the spatial distributions of Au 4*f* and C 1*s* photoelectron emission, respectively. In the Au 4*f* image, the bright region corresponds to the Au-coated substrate, and the dark region shows the graphene-covered areas. This shows good agreement with the OM image in Fig. 1 (a). However, C1*s* chemical mapping did not clearly indicate the SLG and BLG regions because of the carbon contamination covering the sample surface. Combining μ-Raman measurements, the SLG, BLG, and MLG regions can be located easily to perform μ-PES characterization.

Figure 2(c) shows the μ-PES spectra taken on the regions of SLG, BLG and MLG using 480 eV photon energy. All spectra show pronounced asymmetry on the higher binding energy side, which implies that these graphene regions show conducting behaviors as opposed to non-conducting carbon materials, which exhibit symmetric line shape.[13] The C 1*s* spectra were fitted by two main peaks, which describe the graphitic $sp^2$ (G) with a full width at half maximum (FWHM) of about 0.8 eV and contaminant (C) states of graphene.[9,10,14] The G peak mainly reproduces the asymmetric line shape following a Doniach-Sunjic function, which strongly relates to the low-energy electron rearrangement as the core hole creation.[15,16] This asymmetry, measured by asymmetry factor *α*, can be used as a probe of the core-hole screening that depends on the levels of graphene defects and imperfections. The SLG, BLG, and MLG G peaks all have similar low *α* values of 0.08, showing high structural perfection. Additionally, the contaminant C peak accounts for amorphous carbon $sp^3$ bonds showing symmetric line shape. It should be noted that the G peak of SLG (284.4 eV) has a lower binding energy of 0.1 eV than that of BLG (284.5 eV) and MLG (284.5 eV). This is because of the non-equal charge transfer to Au between SLG and BLG, which is dominated by different work functions between SLG (4.6 eV) and BLG (4.7 eV) to Au (5.54 eV).[17], Moreover, BLG and MLG show nearly identical C 1*s* spectra, implying that no charge transfer difference existed between them.

Figure 3 shows the time-dependent C 1*s* core-level photoelectron spectra measured on another SLG area (within the same SLG sheet). Using the same peak decomposition, this SLG area was covered by more contaminant showing a larger C peak contribution in the C 1*s* core-level spectrum. The time-dependent spectra show that the G peak position remained fixed at 284.4 eV throughout the SR illumination. This suggests that the thin Au coating prevents local charging and is extremely suitable for precise core-level PES measurement under long-time signal recording with relatively weak signals. The C/G intensity ratio decreased significantly by about 40 % while taking the PES spectra 5 min (inset of Fig. 3), implying that x-ray illumination-induced desorption of the contaminant on the graphene surface is possible. After 10 min, the C/G intensity ratio decreased gradually, which can be attributed to the nature of the contaminant that exists between the graphene and Au surface is difficult to desorb by x-ray illumination.

In the graphene-metal contact system, work function difference leads to electron transfer between graphene and metal, and naturally makes graphene as *n*-type or *p*-type doping.[4,18] On the conducting Au-SiO$_x$ substrate, the metal-contact-induced doping of graphene can be easily examined and characterized using a core-level PES technique. Being a reference for the graphene doping level, the suspended exfoliated SLG on a partially etched SiO$_x$ substrate (with trench pattern of 3 μm in width and 250 nm in depth) with a Au micro-electrode contact was used as intrinsic graphene for comparison. Figure 4 shows the suspended SLG C 1*s* core-level photoemission spectrum with the same peak decomposition. The contribution of the contaminated C peak in the suspended graphene C 1*s* spectrum is less than that of Au-supported graphene. This suggests that the contribution of contaminated carbon in the core-level spectroscopy of graphene is mainly coming from the supporting substrate. It should be noted that the G peak of SLG on the Au-coated substrate has a binding energy (284.4 eV) lower than the suspended SLG (284.8 eV) of 0.4 eV. Because the measured energy difference between the C 1*s* core level and the Dirac point energy ($E_D$) of graphene is a material constant, that is, independent of environment and substrate-induced effects. Therefore, the core-level shift is directly related to the $E_D$ shift. With the Fermi level ($E_F$) reference on the suspended SLG (which is at $E_D$), the $E_F$ of SLG on the Au-SiO$_x$ substrate was 0.4 eV lower than $E_D$ as p-type doping, which is schematically shown in Fig. 4. This result is reasonable because the SLG has a work function (4.6 eV) lower than Au (5.54 eV), which makes electron transfer from graphene into Au resulting in the p-type doping. And the value of the $E_F$ shift (0.4 eV) is in good agreement with previous calculation result on graphene contacting to Au with a separation of 5 Å.[4] For comparison, the larger BLG work function resulted in less charge-transfer-induced p-type doping, and shows that the BLG C 1s core-level was 0.1 eV higher than that of SLG on Au-SiO$_x$. In addition, based on the linear dispersion of the density of states near the SLG $E_D$, the energy different between $E_F$ and $E_D$ can be used to evaluate hole concentration with (1),

$$N_h = \frac{4\pi}{h^2 v_F^2}(|E_F - E_D|^2) \ , \tag{1}$$

where $N_h$ is the hole concentration and $v_F \approx 10^6$ m/s is the Fermi velocity of graphene.[19] The induced hole concentration of SLG was around $1.1 \times 10^{13}$ cm$^{-2}$ and showed a significant band structure shift, which can be successfully characterized by core-level PES.

In summary, a thin Au film was deposited on SiO$_x$, which is normally used as a graphene-supporting substrate. Because of the elimination of charging effect, graphitic and contaminated carbon states of graphene can be clearly characterized in C 1$s$ core-level spectra. Different levels of Au-coating-induced $p$-type doping on SLG and BLG sheets were also examined in the C 1$s$ core-level shift. Combining thin Au film and SiO$_x$ substrate is a simple but high-throughput method to *in situ* study various hybridization states of graphene by PES or other excited-electron-detection techniques without complicated micro-fabrication processes for metallic contact.

This work was supported by the National Science Council in Taiwan.

**Figure Captions:**

**FIG. 1 (color).** (a) Optical microscopic image of single- (SLG), bi- (BLG) and multi-layer graphene (MLG) sheets exfoliated on Au-coated SiO$_x$ substrate used for μ-Raman and SPEM/S measurements. (b) Comparison of calculated (solid line) and measured (blue dots) contrast of graphene on thin Au ($t_{Au}$=9 nm) coated SiO$_x$ substrate ($t_{SiOx}$=300 nm). (c) μ-Raman 2D band spectra obtained at the areas of SLG, BLG, and MLG. The inset shows that the G band spectra are strongly influenced by Au Raman signal.

**FIG. 2 (color).** SPEM/S measurements of the SLG and BLG sheets on Au-coated SiO$_x$ substrate by using an incoming SR photon energy of 480 eV. (a) (b) Chemical mapping images taken from selected energy channels show the Au 4$f$ and C 1$s$ distinctive chemical regions, respectively. The blue and red dashed lines mark the boundaries of SLG and BLG, respectively, which are consistent with the optical microscopic image shown in Fig. 1(a). (c) μ-PES C 1$s$ core-level spectra taken in the regions of SLG, BLG, and MLG.

**FIG. 3 (color).** Time-dependent C 1$s$ core-level spectra of another SLG area obtained within 20 minutes. The positions of SLG G peaks are constant with SR illumination time, and the C peak intensity relative to that of G peak of SLG is as a function of SR illumination time (shown in the inset).

**FIG. 4 (color).** (a) μ-PES C 1$s$ core-level spectra of the suspended and the Au-supported SLG sheets for determining the Au-induced doping level in graphene. Note that the core-level spectra are aligned to the corresponding core levels at same sample contact Au. (b) Energy-band diagram of intrinsic and Au-induced *p*-type SLG sheets based on the measured values of their C 1s core-level energy difference in (a).

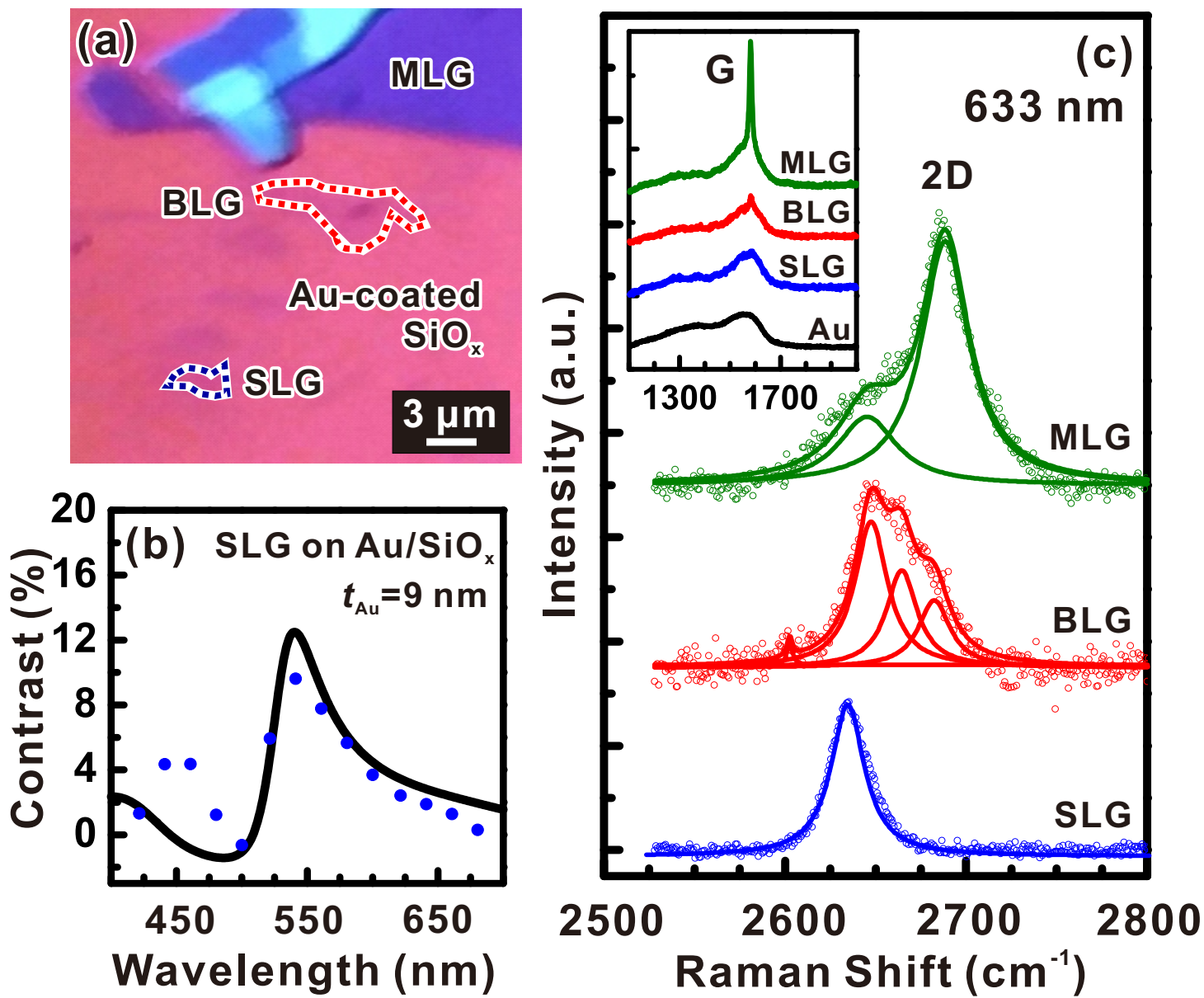

Fig. 1

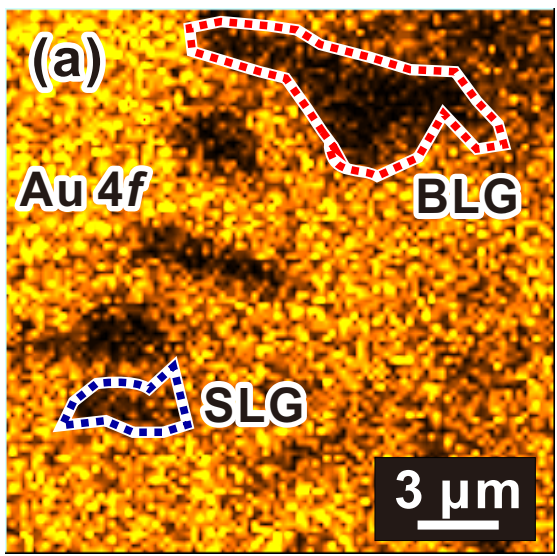
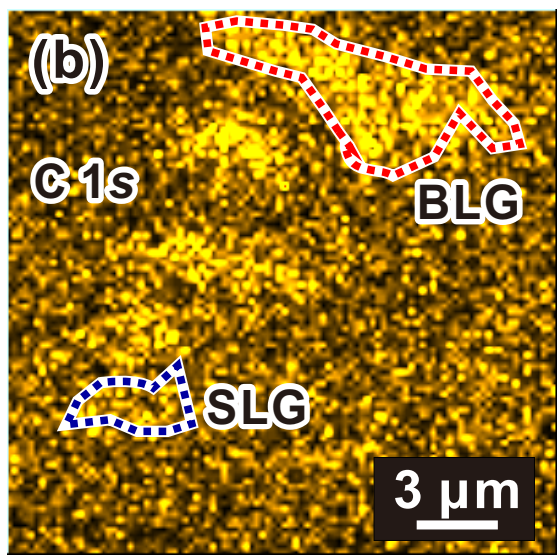
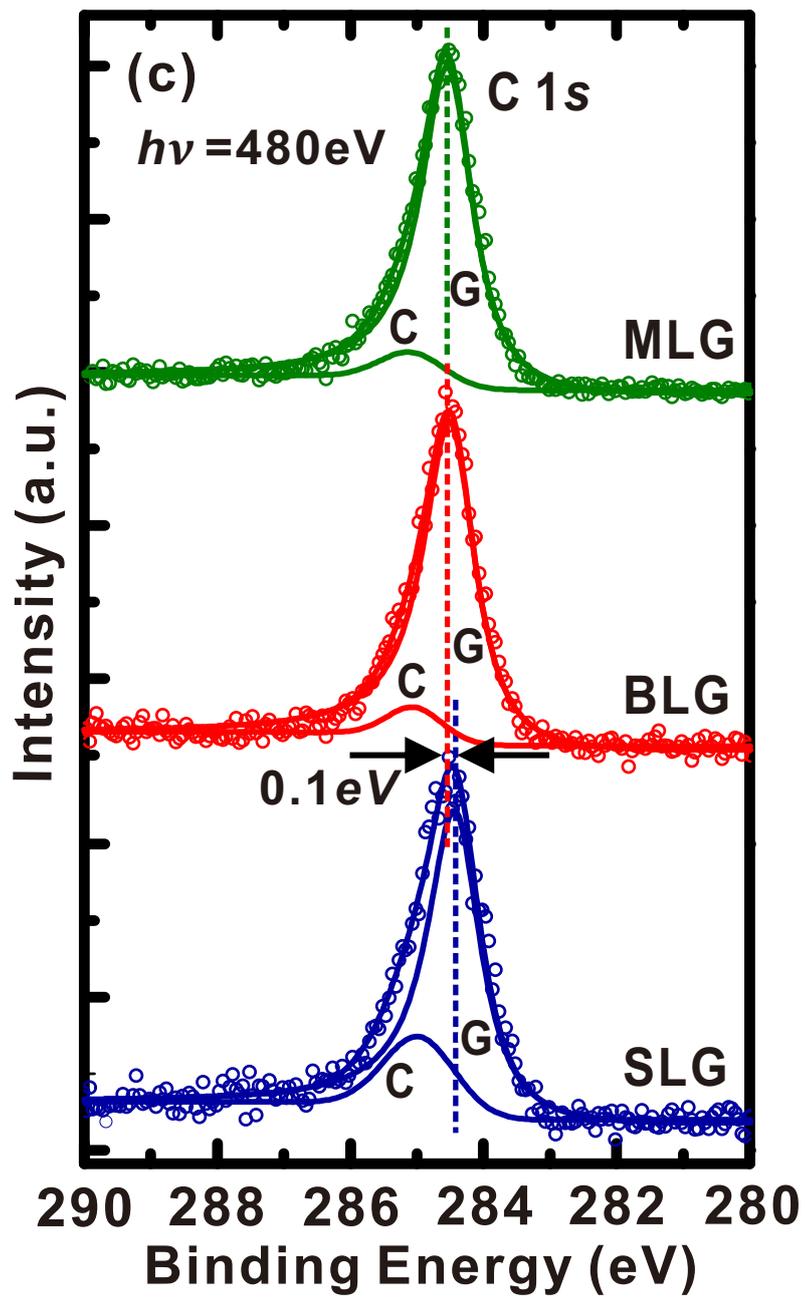

Fig. 2

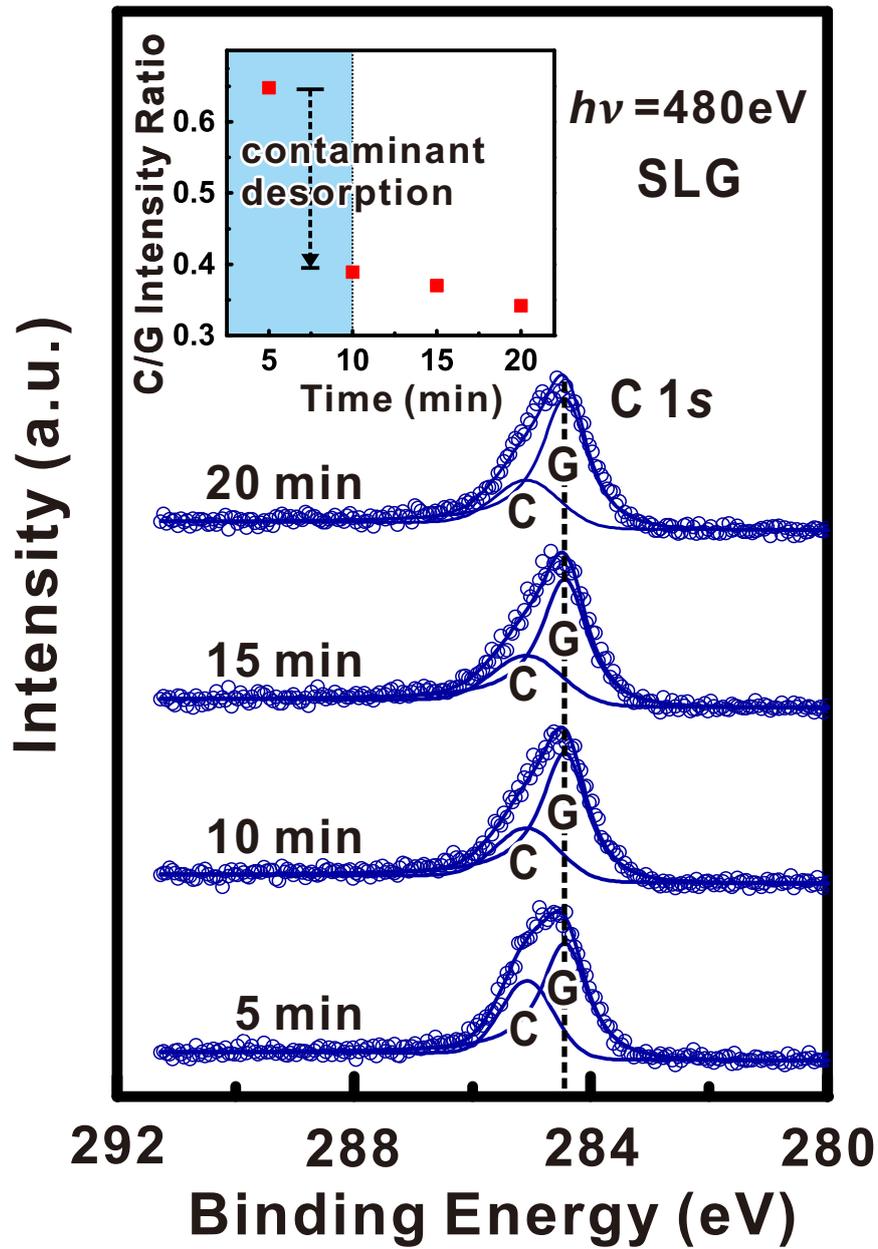

Fig. 3

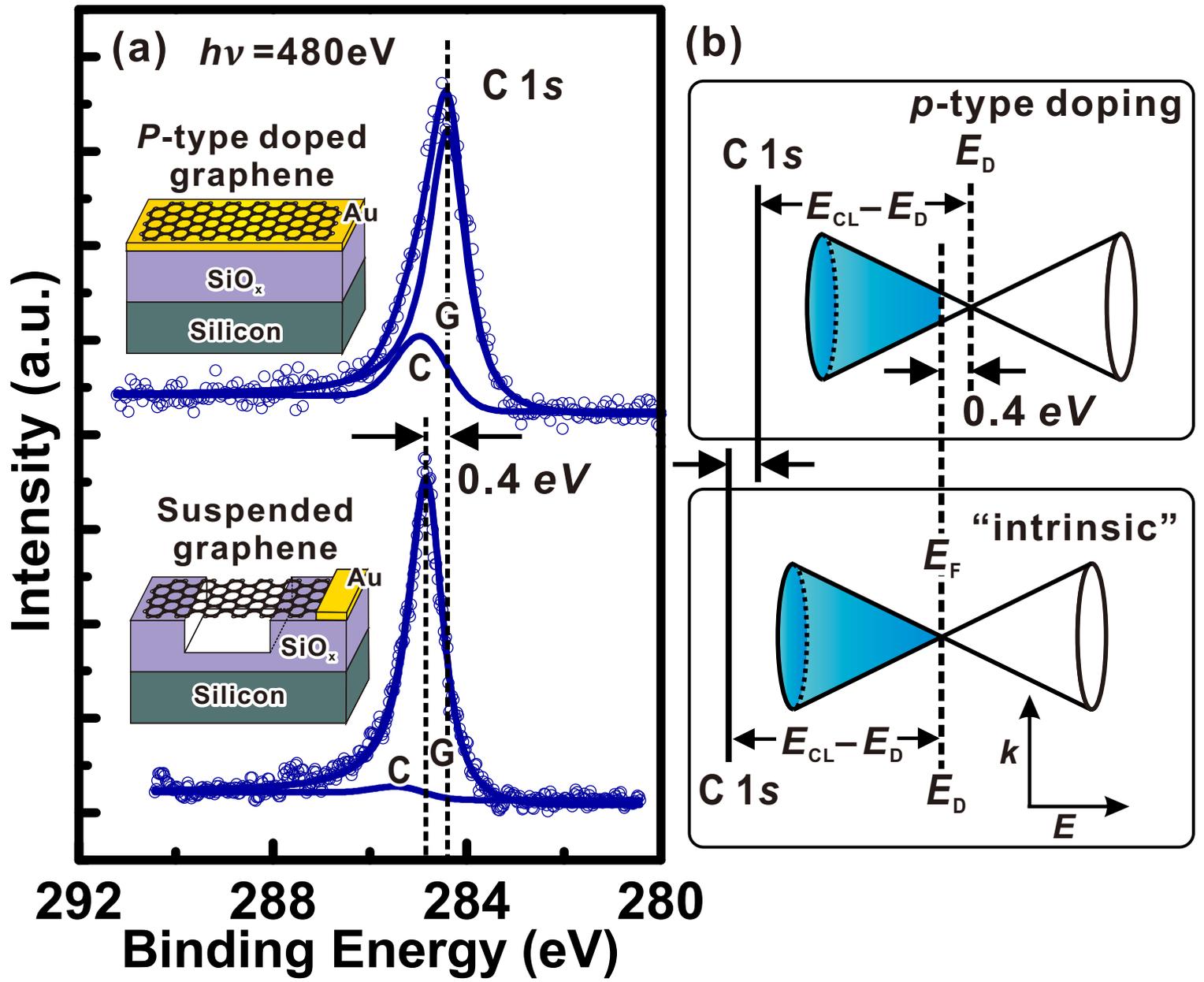

Fig. 4